# Interoperable Provenance Authentication of Broadcast Media using Open Standards-based Metadata, Watermarking and Cryptography

John C. Simmons, Joseph M. Winograd

Verance Corporation, USA

### Abstract

The spread of false and misleading information is receiving significant attention from legislative and regulatory bodies. Consumers place trust in specific sources of information, so a scalable, interoperable method for determining the provenance and authenticity of information is needed. In this paper we analyse the posting of broadcast news content to a social media platform, the role of open standards, the interplay of cryptographic metadata and watermarks when validating provenance, and likely success and failure scenarios. We conclude that the open standards for cryptographically authenticated metadata developed by the Coalition for Provenance and Authenticity (C2PA) and for audio and video watermarking developed by the Advanced Television Systems Committee (ATSC) are well suited to address broadcast provenance. We suggest methods for using these standards for optimal success.

## 1. Introduction

In our interconnected world, information flows ceaselessly, shaping opinions, policies, and societies. Within this digital torrent false and misleading information often obscures the truth.

False information may take the form of misinformation, spread when well-intentioned individuals share what they found online, neglecting to verify what they found. Or it may be disinformation, false or misleading information intentionally created and spread to deceive.

Both forms of false information are harmful, and both thrive in the global digital ecosystem. Social media platforms amplify their reach, turning falsehoods into viral storms. A rumor, a manipulated video, a fabricated statistic—these can cascade across screens, eroding public discourse.

### 1.1. Provenance and Authenticity

Any attempt to address false information on the web must proceed from an understanding of how people come to place trust in information.

The prevalence of information 'bubbles' demonstrates that people primarily place trust in specific sources of information. If information appears unaltered and from a trusted source, we often consider that information to be factual.

In other words, most of us judge what is factual based on the provenance and authenticity of the information, where provenance refers to the origin, history, and chain of custody of a piece of audio-video content, and authenticity refers to whether the content has been manipulated or altered in a way out of the control of the trusted source of the information.

### 1.2. The Role of Standards

There are two general methods for conveying provenance and authenticity metadata in association with audio-video content. Metadata can be cryptographically bound to the audio-video content, perhaps stored at the audio-video container level. Metadata can also be embedded as a watermark in the audio-video elementary stream.

For practical reasons described in this paper these two metadata approaches are interdependent. Both cryptographic and watermarking provenance and authenticity methods will be required to provide a reasonable degree of provenance assurance.

A critical issue to address is the impact of adopting proprietary solutions on interoperability and scalability, an issue often encountered. For example, fifteen years ago, Digital Rights Management (DRM) on the web had not yet been standardized. Prior to the ISO/IEC Common Encryption standard [14] playback devices would have to support every major variety of digital rights management software, and there would be as many versions of the audio-video content as there were DRM systems. Had this continued it would have resulted in a combinatorial explosion, an effective barrier to large scale growth of commercial web media. It is no wonder that Netflix was one of the first companies to recognize the value of the common encryption standard.

It is reasonable to expect that the same will hold true for provenance and authenticity. For scalability and interoperability, the cryptographic metadata bound to the audio-video container and the watermark metadata embedded in the audio-video elementary streams must include open standard options.

### 1.3. Scope and Goals of this Paper

A solution for provenance and authenticity for broadcast content distributed on social media platforms is outlined, utilizing metadata, watermarking and cryptographic standards. Our goal is to show how this can be used with broadcast news content while pointing out several important implementation considerations.

## 2. Provenance and Authenticity Success Scenarios

A provenance and authenticity use case can encompass multiple scenarios, including **success scenarios**, where everything goes roughly as intended and various ***exception scenarios***, which lead to undesirable outcomes. All these scenarios should be describable as discrete programmatic steps to uncover the functional requirements for addressing provenance and authenticity in practice.

In preparing this paper, we examined the details of one specific, provenance and authenticity use case – the posting of what appears to be broadcast news content to a social media platform. We were particularly interested in the interplay between provenance validation using tamper-evident cryptographic bindings and metadata retrieval using elementary stream watermarks, with a focus on the constituent 'success' and 'exception' scenarios.

A broadcaster produces content for linear distribution by an affiliate/network/platform operator. This content consists of a series of audio-video programs comprising a single linear broadcast TV channel.

There are a variety of scenarios where some of the broadcaster content finds its way into Internet distribution and is uploaded to a social media platform. At a minimum it will then be transcoded into a platform's preferred framerates, resolutions, bitrates, codecs, and container formats. It may also be truncated to meet the platform's maximum size limits.

### 2.1. Verifying Authenticity

Before being posted to a social media platform, broadcast news content may be altered such that there are observable, meaningful differences between what was depicted in the original broadcast and the posted video. This manipulation could be done for artistic, creative, or deceptive reasons, depending on the intention of the editor.

One way to characterize these differences is to ask whether the posted video is an authentic representation of the original, whether it is true to the original, without any judgement as to whether the original itself depicted what transpired in front of the camera lens and microphone.

In this definition, an authentic representation of the broadcaster content may not be bit-wise identical to the original, it may be an unaltered clip from the original, or it may be a transcoding of the original, but it may nonetheless accurately represent what was depicted by the original.

The C2PA standard provides a mechanism for editing the original – e.g., transcoding or clipping – and a means to cryptographically verify the authenticity of the result, but this requires that the tool used to alter the broadcaster content supports the use of this standard. We believe it is highly unlikely that in the near-term social media platforms will reject content that was edited using a tool that does not implement cryptographic metadata standards.

## 2.2. Verifying Provenance

When consuming video in a linear TV receiver, consumers quite reasonably believe that the network/platform operator is accurately identifying the channel and content creator.

Video content on the Internet might misrepresent the identity of the original content creator, the identity of who subsequently transcoded the video and what authorized or unauthorized changes were made.

One way to characterize this history is to use the term "provenance," meaning, the identifiable source of the content and an accurate history of the content's transformation from that source.

There are cryptographic methods for verifying the provenance of video posted to a social media platform using the C2PA standards, but again we believe it is unlikely in the short-term that social media platforms will reject videos that do not enable the use of these methods to identify the source of the original content.

## 2.3. Canonical Representation of a Media Object

A tamper-evident cryptographic binding to an audio-video media object which contains provenance information can be used to validate the provenance and authenticity of that object. It is surely a successful outcome if the provenance is validated, but what should happen if the provenance and authenticity fail to be verified? What constitutes success in this scenario?

There are multiple scenarios where the content may have been innocently modified by the user when preparing to post to a social media platform, since even a single bit change to the content will invalidate a cryptographic binding. Generating numerous alerts for innocent alterations to the media could lead to "security alert fatigue," diminishing user trust in the alert's salience. Doing nothing is also an unattractive option because it leaves users blind to the "trust signal" conveyed by the presence of a provenance assertion.

Should the cryptographic verification of the provenance and authenticity of a media object fail, a successful outcome is for the social media platform to use an embedded watermark to retrieve the authoritative version – the canonical representation of the media object. This can be done by first using the watermark to retrieve the cryptographic metadata associated with the original media object as distributed and then use that trusted metadata to retrieve the media object's canonical representation.

## 3. An Approach to Authentication using Metadata and Watermarking

### 3.1. Architecture

The provenance and authenticity approach in this paper builds on the relationship between the registered content distributors, media objects, their embedded watermarks, associated cryptographic metadata, and the canonical representations of the media object itself. This relationship is shown in Figure 1.

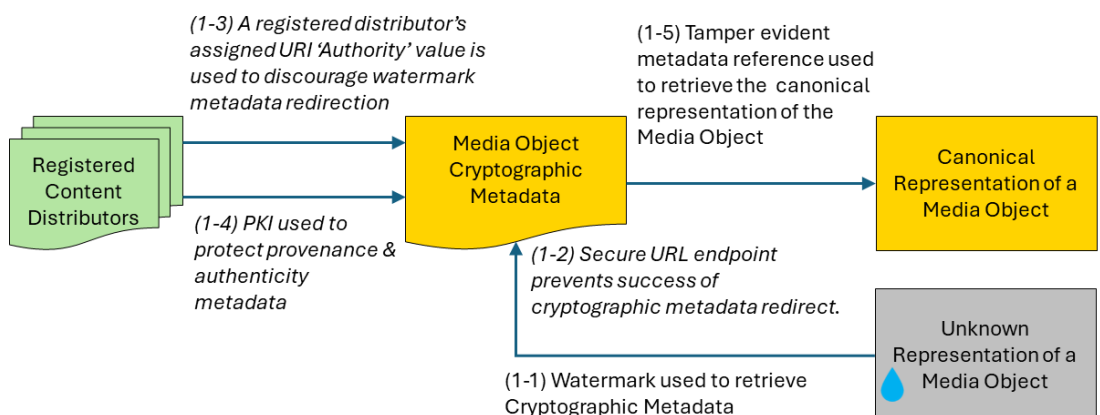

**FIGURE 1 - METADATA AND WATERMARKING ARCHITECTURE**

## 3.2. Security Model

If the tamper evident cryptographic metadata associated with a media object is stored as a component of the media object container, it is relatively easy to remove. A durable embedded watermark can enable cryptographic metadata to be brought back into association with the media object.

Watermark security is typically maintained by making the watermark difficult to remove or alter by keeping the watermark technology secret. This approach works against availability and interoperability by demanding hardened implementations and strict access controls. It can also provide only weak security assurances because its secrecy impedes comprehensive security assessment. Because recorded broadcast content has a long lifespan on the Internet, the security of "closed" watermarks requires successful long-term protection of the associated secrets. And furthermore, recent advances in attacks on watermarking have demonstrated that advances in artificial intelligence render even robust, secret watermarks automatically removable [10], further diminishing their potential advantages.

This motivates a security approach that does not treat the watermark as a root of trust. Instead, we assume that they are *durable*, i.e. that they survive content processing that causes traditional metadata formats to be lost, but that they are otherwise as *mutable* as traditional metadata and can be modified or removed by any intermediary. Like traditional metadata, data conveyed via watermarking is treated as untrusted and must be validated using cryptographic methods.

This same approach was advocated by England et al. [12] in their foundational work on media provenance authentication. That work, however, assumed the presence of a signature in the watermark payload. We view that signature as unnecessary and assume that the watermark carries only a URL and media timeline. The root of trust is a manifest that has been retrieved using the watermark and cryptographically validated using an appropriate trust list.

## 3.3. Watermarking Audio-Video Content

Our success scenario demands a path to validated content regardless of distribution source, which for broadcasters must encompass both linear and on-demand delivery. To achieve this, it must be possible to apply watermarking in asset-based digital publishing as well as within the live production chain.

Figure 2 illustrates an exemplary production flow in which watermarks are applied to enable provenance across multiple distribution paths, with asset watermarks applied to pre-recorded assets and service watermarking continuously applied to a linear playout stream.

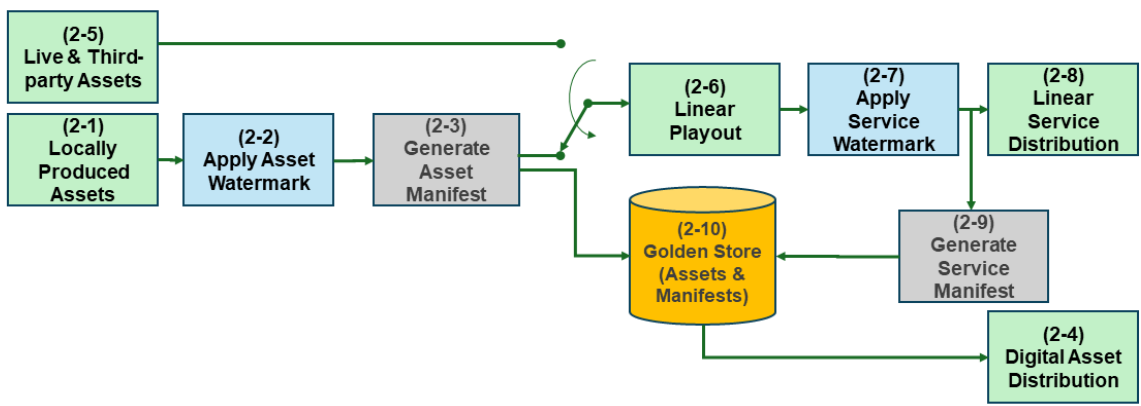

**FIGURE 2 - SERVICE AND ASSET WATERMARKS**

The broadcaster may produce content to be published on their website (2-1). They would apply an asset watermark (2-2), generate cryptographic metadata or a "manifest" for that content (2-3), store the asset (2-10) and distribute the asset to their website (2-4).

The broadcaster may also want to take live and third-party assets (2-5) and prepare them for linear playout (2-6). They would apply a service watermark with a time varying component (2-7), distribute the content (2-8) and periodically generate cryptographic metadata or "manifest" information for that broadcast (2-9).

A watermark can be used to retrieve the associated, static cryptographic metadata and canonical content. And the time-varying service watermarks can be used to retrieve the associated, time-varying cryptographic metadata and canonical content (2-10).

### 3.4. Validating Content Authenticity using Watermarking

Figure 3 and Figure 4 summarize how watermarks can be integrated into the content validation process.

Media validation uses the cryptographic metadata which may be stored in the media object, distributed with the media object and/or retrievable from the cloud. This object is referred to as a 'manifest' in the C2PA standard [4].

### 3.4.1. Media object validation scenarios

If the content can be validated by a contained or retrieved manifest, then a successful outcome does not require utilizing canonical content.

If the media contains a manifest (3-1), the manifest corresponds to a registered distributor (3-2), and the manifest validates the media object (3-3), validation is achieved without reference to a watermark. We would view this as a **success scenario**.

Otherwise, if the media object does not contain a watermark (3-4), the media remains unvalidated, this is an *exception scenario*.

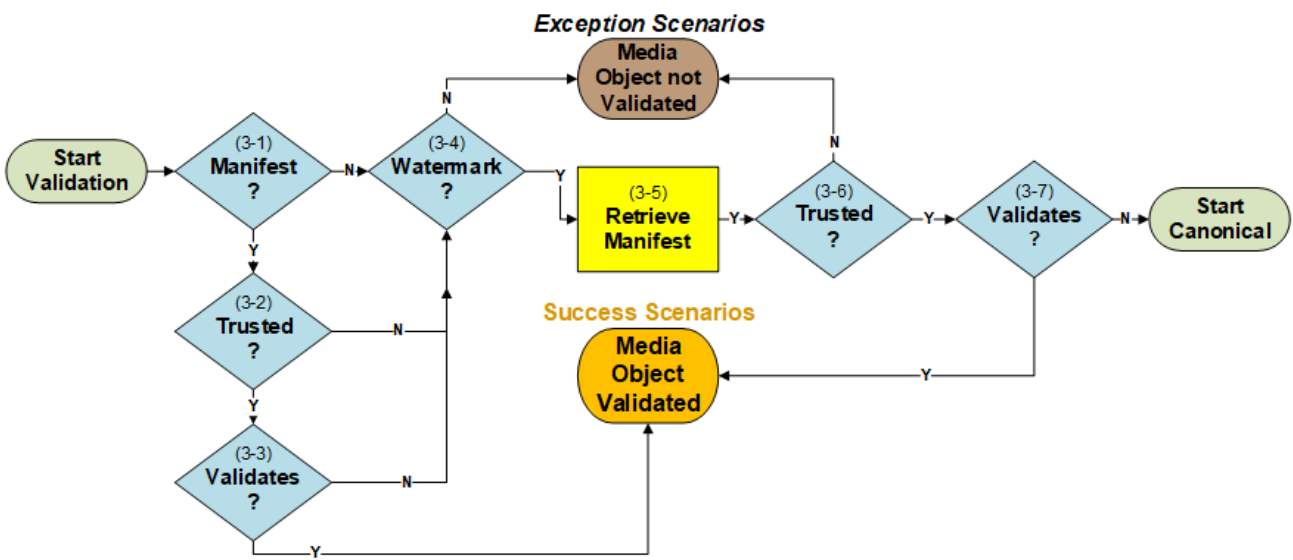

**FIGURE 3 - MEDIA OBJECT VALIDATION SCENARIOS**

If the media does contain a watermark (3-4) then the manifest is retrieved from the manifest cloud store (3-5). If this retrieval fails, for example if the URI Authority field provided in the watermark does not correspond to a registered broadcaster, the media object is not validated, an *exception scenario*.

If the retrieved manifest's digital signature does not correspond to an approved broadcaster (3-6), then the media object cannot be validated. Another *exception scenario*.

Otherwise, if the manifest's digital signature is trustworthy (3-6) and the manifest validates the content (3-7), validation is achieved by using the watermark. A **success scenario**.

### 3.4.2.  Media object canonical representation scenarios

If a retrieved manifest (3-5) is trustworthy (3-6) but it does not validate the content (3-7), then it is the view of this paper that the only **success scenarios** involve canonical processing.

The decision to perform canonical processing (4-8) can be made by the user posting the content or by the platform supporting the validation logic, depending on the policy being adhered to by the social media platform.

If the decision is to perform canonical process (4-9), the validator retrieves the canonical content (4-10).

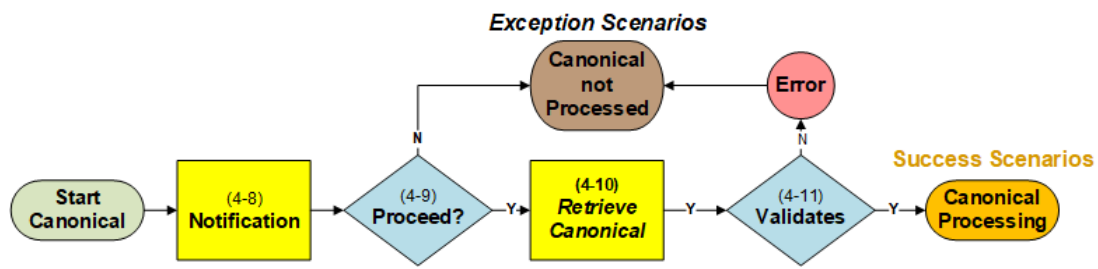

**FIGURE 4 - MEDIA OBJECT CANONICAL PROCESSING SCENARIOS**

The previously retrieved manifest (3-5) should always validate the canonical content (4-11). If it does not, it is an error and an ***exception scenario***.

We view validation of the retrieved canonical content as optional because its retrieval location has been established as trusted through validation of the manifest that contains it (3-6) (3-7).

### 3.4.3.  Media object canonical processing

The availability of a canonical version of the media object presents the social media platform with additional **success scenario** opportunities, including one or more of the following:

- Posting the uploaded content together with the retrieved asset, or a link to it.
- Providing the uploader with a choice between which version of the content should be posted and posting that version with an appropriate label.
- Performing an automated comparison of the uploaded and reference asset to determine the nature and amount of difference between the two.
- Automatically replacing the uploaded content with the valid asset content.
- Forwarding the uploaded content and the retrieved asset to an internal content moderation process.

## 4.  This Approach Applied to Live Broadcast

### 4.1. Low Latency Considerations

Real-time broadcast and live streaming, often referred to as "glass to glass," is a process where content is captured through a camera lens and transmitted to a viewer's screen with minimal delay. Although it is a real-time transmission, it always involves some degree of delay or latency, incidental and/or intentional.

Live scenarios may be categorized by the degree of latency required. This depends on the content's nature and the desired viewer experience. Real-time, low latency live is essential for live sports and breaking news, where timely viewing is important. Higher latency can be introduced for any number of reasons. For example, content is often recorded, edited, or processed before broadcast.

Using digital signatures to protect provenance metadata for 'glass to glass' real-time live streaming scenarios is technically challenging. The primary difficulty is that performing a digital signing operation on a Content Delivery Network (CDN) edge server is not adequately secure and performing that operation in a Hardware Security Module (HSM) is unlikely to achieve the low latency desired.

However, any live scenario where the content is captured downstream, edited, and subsequently posted will introduce an inherent latency sufficient to allow the use of an HSM for provenance metadata protection.

## 4.2. Live Broadcast News Content Posted to a Social Media Platform

A 30-minute evening news program is broadcast. The live broadcast is captured and recorded on a device downstream of an HDMI port. A 20-second clip of the news broadcast is created as an MP4 file and posted to a social media platform.

No manifest can be present with the content since only the elementary stream will make its way beyond the HDMI port. The social media platform can examine the posted video for a watermark, but what would be the success scenario?

### 4.2.1.   The fragmented MP4 broadcast replica

We believe that the following approach can provide a reasonable degree of provenance and authenticity assurance for broadcast news content posted to social media platforms.

The live news program is broadcast with a watermark consisting of a constant service/asset identifier component and a time-varying index code (Figure 5). This watermark approach is in use today for delivering metadata for interactive television services [1][2][3] and can readily support the retrieval of provenance metadata without the need to modify the watermark itself.

| Watermark Constant | Service/Asset Identifier | | | | | | | | | |
|---|---|---|---|---|---|---|---|---|---|---|
| | Data Hash Segment #M | | | | | Data Hash Segment #N | | | | |
| | Cryptographic Metadata, C2PA Manifest #M | | | | | Cryptographic Metadata, C2PA Manifest #N | | | | |
| | fMP4 Replica, range #M | | | | | fMP4 Replica, range #N | | | | |
| | begin time (M) | +1 Δt | +2 Δt | ... | end time (M) | begin time (N) | +1 Δt | +2 Δt | ... | end time (M) |
| Watermark Time code | BINX(M) | +1 | +2 | ... | EINX(M) | BINX(N) | +1 | +2 | ... | EINX(N) |

FIGURE 5 - DATA HASH SEGMENTS, CRYPTOGRAPHIC METADATA AND WATERMARKS.

The broadcaster or the network/platform operator on behalf of the broadcaster produces a secure transcoding of specified portions of the live linear broadcast into a fragmented MP4 format – an 'fMP4 Replica' of the portion of the linear broadcast for which provenance and authenticity is to be established.

Periodically a C2PA manifest is produced for this fMP4 Replica. In this design the portion of the Replica that each manifest corresponds to is defined as the Data Hash Segment (DHS). The real-time duration of a DHS defines a minimum lag time behind the linear live edge for the availability of DHS Replica Manifests.

The Replica itself consists of a sequence of fragmented MP4 segments or chunks for each track, adequate to cover the length of the Data Hash Segment. Each segment or chunk includes auxiliary 'c2pa' boxes [4] which can be used by a C2PA validator to validate any portion of the DHS Replica, as described below.

Apart from the addition of c2pa-specific ISOBMFF boxes, the Replica format is identical to the format in common use for adaptive bitrate streaming, the Common Media Application Format or CMAF [15].

### 4.2.2. The fragmented MP4 replica C2PA manifest

The fMP4 Replica Manifest is constructed in the exact same way as a C2PA Manifest for audio-video streaming [section 9.2.3, 4].

Before the manifest is generated, a DHS initialization segment is produced for the content stored in the DHS Replica. The cryptographic metadata stored in this initialization segment is identical to that specified by C2PA for adaptive bitrate delivery [9.2.3, 4].

The c2pa-specific box in each track's initialization segment will contain the C2PA manifest, which, as is the case for adaptive delivery, must be identical across tracks. The Manifest's `c2pa.bmff.hash` assertion will contain CBOR with an array of merkle rows, one per track.

In the C2PA specification for adaptive delivery provenance validation, the Merkle tree associated with the entire video stream enables piecewise validation of individual fragment components of the stream without access to the entire stream [4]. The same mechanism enables piecewise validation of arbitrary portions of the DHS using a single DHS manifest.

### 4.2.3. Producing a canonical live recording

If the watermark in the live recording is time-varying, it can be used create a canonical live recording, as shown in Figure 6.

The time-varying watermark is used to derive the manifest recovery URL. OTT BINX and OTT EINX are the time indices corresponding to the start and end of the posted video, respectively.

A recovery request (6-1) is sent. The DHS Manifest is provided in a recovery response (6-2).

This recovery request response is identical to the method used today for interactive television. The only change is in the payload of the response from the provenance-authenticity server.

The DHS corresponding to the manifest is accessed from the Asset Reference Assertion in the DHS Manifest (6-3).

The fMP4 Replica is used as an authenticated mezzanine format, to produce a canonical MP4 representation of the posted live content. Any portion of the Data Hash Segment can be validated with the DHS Manifest.

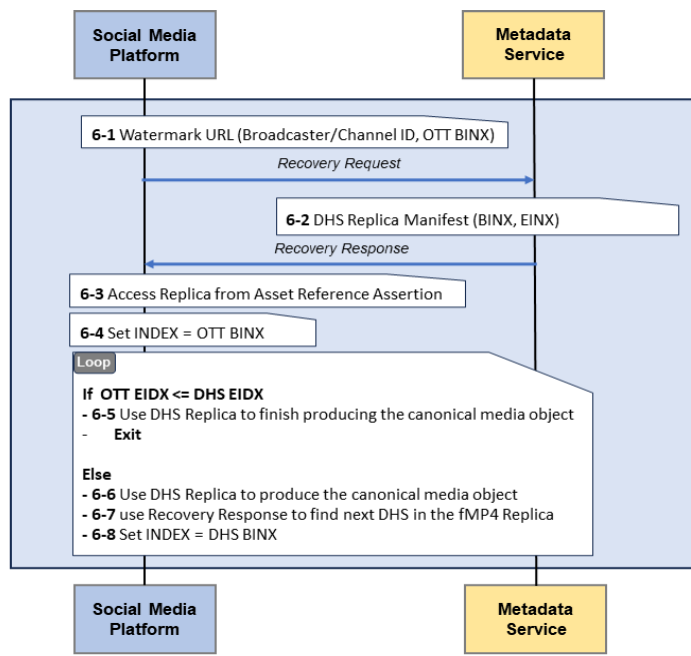

**FIGURE 6 – PRODUCE A CANONICAL MEDIA OBJECT**

Beginning with the OTT BINX (6-4), the algorithm walks the Data Hash Segments provided in the Recovery Response (6-7) until the EIDX of the posted live recording is reached (6-5).

## 5. This Approach Applied to Web Published Content

### 5.1. Differences without a Distinction

During validation of content posted to a social media platform, even the slightest alteration to the content can cause it to be flagged as inauthentic. There are multiple scenarios where the content may have been innocently modified by a user, making their edits from a provenance perspective a 'difference without a distinction'.

As discussed, we believe the successful outcome for a validation failure to be for the social media platform to recover the original content and use it in one of the ways we outlined. There are cases, however, where this too will result in an unsuccessful outcome.

## 5.2. Clipped Web Published News Content Posted to Social Media

One of the most likely such cases is what we are calling "the clipped news segment" scenario.

Consider the following example. The broadcaster publishes a 30-minute evening news program to their website. The published video file includes cryptographic metadata, and it is watermarked. A user wishes to share a 20-second clip from that 30-minute program. They download the broadcaster published video file, edit it to produce a 20-second clip, and attempt to post it to a social media platform.

If the editing tool the user used removed the manifest, the social media platform can recover the metadata using the watermark, as described above. Regardless, the file will be flagged as inauthentic. And recovering the canonical version of the content will result in a 30-minute post.

### 5.2.1. Producing the canonical news clip

If the watermark in the clipped news content is time-varying, it is used to derive the manifest recovery URL, a recovery request (1) is sent where BINX is the time index corresponding to the start of the clip. The retrieved DHS Manifest in a recovery response (2) can be used to produce a canonical version of the news clip, as shown in Figure 6, following the same steps as producing a canonical live recording.

Since social media platforms transcode posted video into a multitude of targeted formats, it is likely that they would treat the DHS as a canonical mezzanine format to produce a wide variety of device targeted formats. Using fragmented MP4 as a mezzanine format is commonly done.[1]

## 6. STANDARDS-BASED SOLUTIONS

A fundamental element of the provenance challenge is that the global Internet facilitates content following an unconstrained path from its place of creation to that of presentation. This makes interoperability between the systems that generate and validate authenticity signals essential.

The use of incompatible technologies creates problems at both ends of the system. If different content producers choose to label their content using incompatible authentication technologies, a platform that could receive uploads of content originating from any of these producers will need to support all the various technologies that they use.  Conversely, if different platforms choose to validate content using incompatible technologies, a content producer whose media could be uploaded to any of these platforms would need to be able to generate authentication signals using all of the platforms' technologies in their content.

The establishment of open technical standards for provenance authentication presents a promising path towards addressing these problems. In the interest of advancing progress in this area, we have considered two candidate standards applicable to broadcast content – one cryptographic, the other watermarking. For each, we assess their suitability in the context of the broadcast use cases described above and, where necessary, identify improvements.

## 6.1. The Coalition for Content Provenance and Authenticity (C2PA) Standards

At the January 2019 World Economic Forum in Davos the director of a major news organization asked Eric Horvitz, Chief Scientific Officer at Microsoft, what could be done about deepfake videos. Within a year of that conversation Microsoft had initiated the Authentication of Media via Provenance (AMP) project, the New York Times had launched the News Provenance Project (NPP) and the BBC and CBC together created the BBC/CBC Provenance Project.

---

[1] In addition, the MPEG DASH specification [Annex C.4, 25] provides support to access segments of presentations at a specified media time through the use of an MPD Anchor, using a query parameter "t=" that a client can append to an MPD URL with either an NPT or UTC time. This could be used to access portions of the fMP4 Replica.

The convening of these parties took place in London, May 2019, resulted in a combined activity - the Origin Project. Also in 2019 Adobe, the New York Times and Twitter founded the Content Authenticity Initiative (CAI), with the stated goal to build a system to provide provenance and history for digital media.

Realizing there was a need for a single, scalable interoperable standard for provenance and authenticity, on February 22, 2021, Microsoft and BBC teamed up with Adobe, Arm, Intel and Truepic to create the Coalition for Content Provenance and Authenticity (C2PA), combining the specification efforts of Adobe, which had focused primarily on still images, and Microsoft, who had focused almost entirely on video assets.

Today C2PA includes 120 companies and continues to grow. The C2PA Steering committee consists of Adobe, BBC, Google, Intel, Microsoft, OpenAI, Sony and Truepic. Members include Amazon Web Services, ARM, Canon, Leica, New York Times, Nikon, and NHK.

### 6.2. C2PA Standards Overview

Because Adobe and Microsoft had independently made considerable progress developing specifications and prototypes before 2021, the C2PA, version 1.0 specification was released to the public in less than one year - January 26, 2022.

At the time of this writing, the most recent version is 2.0, released January 2024 [4].

#### 6.2.1. Normative and informative C2PA documentation

There are presently two C2PA technical specifications, 1) "Content Credentials: C2PA Technical Specification" and 2) "Attestations in the C2PA Framework" [4].

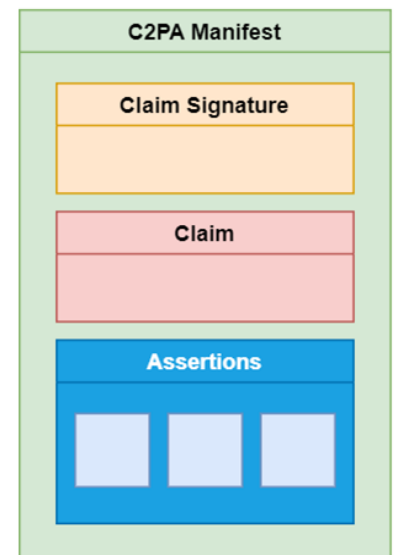

FIGURE 7 - C2PA MANIFEST [4]

Released along with these normative specifications are a collection of informative documents, including an explainer [16], guidance for implementers [17], user experience guidance [18], security considerations [19], harms modelling [20] and guidance for artificial intelligence and machine learning [21].

The following overview introduces a few of the central concepts in the C2PA specifications, skipping over topics not central to this paper. Details can be found in the specifications themselves.

#### 6.2.2. Content credentials overview

C2PA metadata for an asset conveys **assertions** such as asset metadata, actions performed, thumbnails, and cryptographic bindings to the content. These assertions convey the provenance of the asset. Assertions are combined with additional information to create a **claim**. The set of assertions referenced by a claim are collected into a logical construct referred to as an **assertion store**. The claim is digitally signed, creating the **claim signature**.

Assertions, Claims, and Claim Signatures and some additional information are combined to form the **C2PA Manifest**. For some formats, the C2PA Manifest may be embedded in the content. (see Figure 7).

For each manifest there is a single assertion store. However, multiple manifests can be associated with an asset, each one representing a specific series of assertions.

#### 6.2.3. Hard and soft bindings

There are two types of bindings supported by C2PA – hard bindings and soft bindings. These are described as hard binding assertions or soft binding assertions in the assertion store.

The hard binding uses a cryptographic hashing algorithm over some or all of the bytes of an asset. It can be used to detect tampering.

The hard binding of an ISO Base Media File Format (ISO BMFF) formatted asset is described in section 9.2.3 of the Content Credentials specification [4]. There is one binding for a monolithic MP4 file where the mdat box is validated as a unit, and a different binding for when the asset is a monolithic MP4 file where the mdat box is validated piecemeal, or when the asset is a fragmented MP4 file. In this paper we reference both the monolithic MP4 and the fragmented MP4 bindings.

Soft bindings may be a perceptual hash computed from the digital content, or a watermark embedded in the digital content. In this paper we describe a soft binding using an open standard watermark.

### 6.2.4.    Manifest store

C2PA data is serialized into a JUMBF-compatible box structure (22). The outermost box is the C2PA Manifest Store, also known as the Content Credentials. How the Content Credentials, the constituent Manifests and assertion stores utilize the JUMBF-box structure is described in section 11.1 of [4].

### 6.2.5.    Update manifests

Most C2PA manifests are standard manifests, containing exactly one hard binding to the associated asset. To accommodate provenance workflows where additional assertions are provided but the digital content is not changed is done using an Update Manifest (see 11.2.3 of [4]).

### 6.2.6.    Embedding manifests

C2PA specifies how to embed manifests to a wide variety of formats, including JPEG, PNG, SVG, FLAC, MP3, GIF, DNG, TIFF, WAV, BMF, AVI, WebP, RIFF, BMFF as well as fonts (see 11.3 of [4]).

### 6.2.7.    Asset reference assertion

This assertion indicates one or more locations where a copy of the asset may be obtained. The location is expressed as a URI. This paper makes extensive use of this assertion to enable canonical processing of watermarked content which fails provenance and authenticity validation (see 18.15 [4]).

## 6.3. C2PA Standards Suitability

### 6.3.1.    Open standard for provenance and authenticity

C2PA provides an open standard to associate a cryptographic binding to monolithic MP4 files with a single or multiple mdat boxes as well as to fragmented MP4 content used with adaptive streaming. It supports a tamper evident manifest store so that the transcoding history of a media object can be provenance ensured. And it provides a mechanism through soft bindings to recover the cryptographic metadata associated with a media object.

### 6.3.2.    Gap analysis

A central premise of this paper is that canonical content processing should occur when the provenance and authenticity of an asset cannot be validated, and a watermark is present.  C2PA supports an asset reference assertion, but at the time of this writing does not provide normative language on how this reference should be used should validation fail.

The C2PA specification deals principally with an asset which either contains a manifest or at one time contained a manifest. This paper suggests that for the case of broadcast content, it is important to define a success scenario for when a watermark is present, but the asset was never published to the web associated with an asset watermark.

These are implementation issues and would not impact interoperability.

## 6.4. ATSC Watermarking

In 2016, the US-based digital television broadcast standards organization ATSC, published standards for use of watermarking technology [1][2][3] in connection with their development of the ATSC 3.0 ("NextGen TV") system.

These standards provide open specifications for the use of watermarking technology and associated network protocols to deliver arbitrary timed metadata associated with media content to network-connected clients through distribution paths that include media processing (e.g. transcoding) and metadata removal (e.g. HDMI, analog reconversion).

ATSC's primary motivating use case for watermarking is enabling access to NextGen TV interactive (two-way) services for viewers who have purchased compatible TVs but who continue to receive broadcast services from other distribution paths, such as STBs, streaming media players, or ATSC 1.0 transmissions. These standards have been commercially deployed in the United States by multiple broadcasters and television equipment manufacturers.

The technology has also been found to be suitable for use with other broadcast systems. Since 2020, the HbbTV and DVB have published a series of standards that employ ATSC watermarking to enable interactivity and targeted advertising in their platforms. These standards are currently being readied for commercial deployment in Germany.

### 6.4.1.    Description

The ATSC watermark system is specified in the publicly available standards ATSC A/334 [1], A/335 [2], and A/336 [3]. Its function is to deliver arbitrary timed metadata using audio and/or video watermarks embedded into media essence. It supports methods for conveying metadata directly in watermark messages or indirectly, by reference, via carriage of a time-tagged URL that identifies a network resource containing the metadata. The architecture of the ATSC watermark system can be understood using the OSI abstraction model shown in Figure 8.

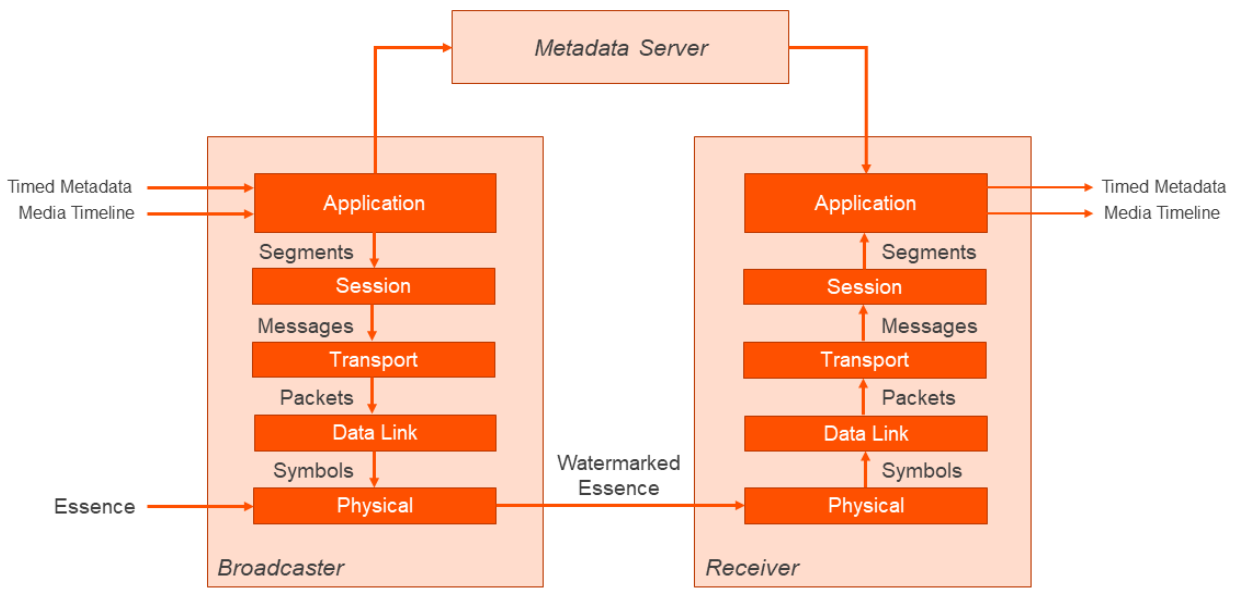

**FIGURE 8 - OSI ABSTRACTION MODEL FOR ATSC WATERMARKING.**

Taken from bottom to top, *essence* is the baseband audio or video signal components. The physical layer consists of a stream of raw binary symbols conveyed as audio and video watermarks in the essence. Audio watermarks are conveyed using autocorrelation modulation in the 2.5k-5kHz band. Video watermarks are conveyed using luma modulation in the top two lines of active video. While watermark insertion and detection are performed on baseband (decoded) essence, the system is compatible with a wide range of media processing algorithms applied to watermarked essence, such as low bit-rate coding, that are typically found in digital media distribution. Both audio and video watermark physical layers also permit watermark energy to be adapted to the content to preserve perceptual quality. In formal testing conducted by ATSC technical committees, the audio watermark was demonstrated to be capable of surviving HE-AACv2 encoding at 32 kbps stereo without performance loss while preserving perceptual transparency. The video watermark was demonstrated to be capable of surviving AVC encoding at 2.5 Mbps 1080p/30.[2]

The data link layer differs for audio and video watermarks, with the audio watermark carrying a sequence of data cells of 1.5 seconds duration, each carrying a 50-bit data packet along with a synchronization header and BCH error protection. The video watermark data link layer conveys a 168-bit data packet in each video frame along with a synchronization header, message framing, and CRC error protection.

The transport layer also differs for audio and video watermarks. For the audio watermark, the transport layer conveys a single packet format, the *VP1 payload*, that conveys a "tiny URL" encoded into two fields – a *server code* that identifies a network server and an *interval code* that identifies a metadata resource on that server associated with the location in the media content where the watermark is embedded. For the video watermark, the transport layer can convey metadata by reference using the VP1 payload or directly, using a variety of messages associated with known broadcast metadata types such as stream events, presentation timestamps, and content identifiers. Server codes are assigned values for which ATSC maintains registry authority.

The session layer specifies constraints on the arrangement of watermark messages within assets that enable receivers to perform reliable decisioning regarding the arrangement of watermarked content, including when a particular watermarked asset starts and ends and where on the media timeline a given media sample lies. This is particularly important in contexts the content arriving at the receiver has been composed from multiple different sources, such as a "mash-up" of multiple sources or edited version of content.

The VP1 payload is relied on in the session layer to provide the context boundary for a media asset. A watermark media asset (which can be either an individual program item or a continuous program stream) carries *audio or video watermark segments* comprised of contiguous, watermarked 1.5 second content intervals with a constant server code value and incrementing interval code values.

At the application layer, directly conveyed metadata becomes valid at the location in the content where it is placed. For metadata delivery over a network, a RESTful application layer protocol between the receiver and a metadata server is specified wherein receivers retrieve arbitrary timed metadata using VP1 payload data. This protocol was adapted by ATSC from an existing 3GPP MBMS protocol [6]. In it, receivers request a metadata resource using a URL constructed from the first VP1 payload that they encounter in a watermark segment. The *authority* portion of the URL is an Internet hostname determined using DNS resolution of the *server code* within a second level domain specified by the standard. The *path* portion of the URL includes the *interval code* within a predefined template. The response is a multipart/related MIME object [23] containing some protocol-specific metadata objects and some number of additional metadata objects. The protocol-specific metadata includes a mapping of the VP1 interval code value onto a media timeline, boundaries on the media timeline for which each of the additional metadata objects is valid, and guidance to receivers on where and when updates to the metadata objects should be requested. Receivers request subsequent metadata updates only as necessary, in correspondence with the validity periods of metadata objects that they have received and the time periods of watermarked segments of the media timeline of content that they process.

Any IANA-registered media type [24] can be provided as an additional metadata object. Receivers are expected to route these objects to application-specific handlers and ignore media types that they do not support.

---

[2] Formal perceptual quality evaluation was not performed for the video watermark technology.

### 6.4.2.    Suitability

The ATSC watermark system provides a number of technical capabilities important to the provenance authentication use case.

#### 6.4.2.1.    Open Architecture

The ATSC watermark system shares the same underlying architecture as the modern Internet. The system stack is based on publicly available specifications that enable independent development of interoperable implementations of all components. It uses federated DNS namespace management governed by ATSC, a not for-profit, internationally recognized standards development organization. And it does not rely on any siloed or proprietary services, freeing broadcasters to host metadata services on servers of their choosing with the ability to transition to new hosts at will. A typical use case illustrating an open metadata retrieval architecture is provided in Figure 9.

#### 6.4.2.2.    Durability

Formal testing overseen by ATSC during the standards development process and subsequent commercial deployments have demonstrated that the audio and video watermarks are reliably recoverable through typical content uses. For audio, this includes common transcoding, downmixing, dynamic range compression, equalization, and analog reconversion processing. For video this includes common transcoding, frame rate conversion, resolution conversion, image enhancement, and analog conversion processing.

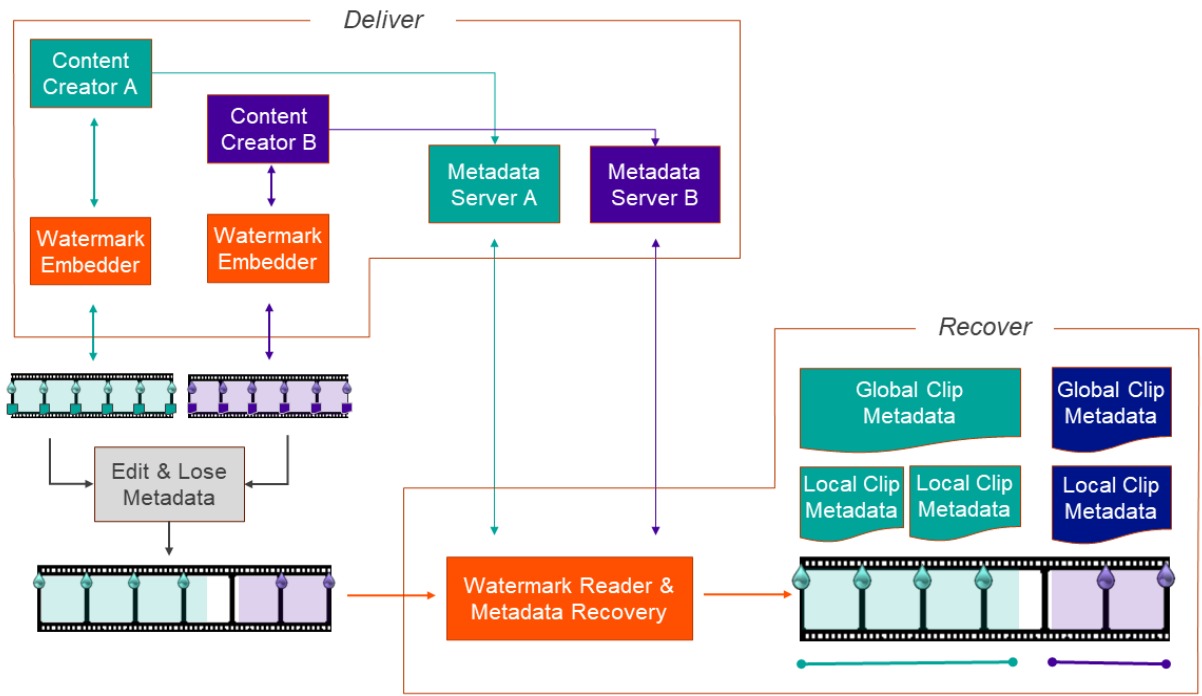

**FIGURE 9 – TYPICAL OPEN ATSC METADATA RETRIEVAL ARCHITECTURE**

#### 6.4.2.3.    Transparency

Formal testing overseen by ATSC during the standards development process demonstrated perceptual transparency of the audio watermark.[3] While formal evaluation of the video watermark has not been performed, the technology has been found acceptable in commercial broadcast settings and representative content is available for public review [7].

---

[3] Independent testing using ITU-R BS.1116-3 [8] methodology found critical samples of audio watermarked content statistically indistinguishable from original unwatermarked content to trained listeners in reference listening conditions.

### 6.4.2.4.   Data Capacity

The approach of carrying time-based references to metadata objects on servers avoids limitation of the size of associated metadata to what can be conveyed by the watermark.

One benefit of this approach is that it allows multiple metadata objects to be associated with content, enabling provenance authentication metadata (e.g. C2PA manifests) to be retrieved using the established protocols for delivering broadcast television signaling (e.g. interactive application data).

Another benefit is that it enables late-binding of metadata to media, which is valuable for live broadcast scenarios, and the ability to revise the metadata associated with content already in distribution simply by posting updated resources to metadata servers, for example to update signatures following a certificate or key revocation.

### 6.4.2.5.   Precision

The latency on the content media timeline to recovery of the first watermark packet with data sufficient to enable recovery of metadata from a network server is 1.5 seconds (minimum) and 2.25 seconds (average) for the audio watermark and 1 video frame (minimum) and 83 milliseconds (average) for the video watermark.

 After initial metadata recovery, session layer logic enables the media timeline to be tracked within the segment with accuracy of ±2 milliseconds from the audio watermark. Media timeline accuracy for the video watermark depends on whether and how message multiplexing is being used but, in any case, it will be between frame accurate and ±83 milliseconds. Segment (i.e. edit) boundary decisioning can be determined ±1 second.

The ability to precisely recover the media timeline and edit points of a segment is an essential capability for provenance authentication of broadcast content. When broadcast content is redistributed, e.g. on social media platforms, it will of necessity be just a clip of the linear program stream. The ability of the watermark technology to determine the clip boundaries with precision enables the client to access metadata associated specifically with that region of the media timeline, align manifest data for validation, resolve asset references to corresponding time segments of valid assets, and support any of the use cases described above (e.g. side-by-side review, automated analysis, content substitution, etc.).

### 6.4.2.6.   Mutability

While ATSC audio and video watermarks are durable, in that they survive content processing that causes traditional metadata formats to be lost, they otherwise have the same mutability properties as traditional metadata, which is that they can be modified or removed as needed.[4]

The mutability characteristic provides a benefit in scenarios where watermarked content is repurposed, such as when an editorial segment includes a news clip that includes syndicated footage, and it is desired that only the provenance information associated with the combined work be referenced by watermarking. In this case, each successive actor can update the watermark to reference a manifest that includes comprehensive provenance information for their output This approach aligns with the C2PA approach to manifest metadata updates, wherein revisions are consolidated within a unitary store [4] rather than simply appended.

## 7.  Conclusions

The trust placed by the public in broadcasters as an authoritative source of information provides disinformation agents with a strong incentive to falsely adopt the broadcaster's imprimatur. Given that powerful generative AI models have been open-sourced, reliance on safety mechanisms being implemented by the generative AI systems against these risks cannot be relied on as a meaningful countermeasure.

---

[4] See, for example, ATSC A/339: Audio Watermark Modification and Erasure [9].

There is a critical need to identify open standards that enable media creators, distributors, and consumers to be able to authenticate the provenance of content securely and reliably.

In this paper we described a broadcaster authentication approach based on the application of open, interoperable, standards-based technologies. Watermarking standards like that defined by ATSC [1][2][3] already in use by broadcasters enable automatic identification of content, media timeline, and provenance metadata by platforms during content ingest or presentation. And provenance metadata standards like that defined by C2PA [4] provide cryptographically-assured validation of authenticity, as well as access to authoritative versions of content.

In our analysis we examined one particular use case in detail, the posting to a social media platform of what appears to be broadcast content. We found that an architecture based on these open standards provides a workable framework for broadcast authentication. We also identified methods to minimize the impact of authentication failures by using these standards to access the authoritative version of the content, even when that version was never itself published on the Internet.

While the challenges addressed in the paper are motivated by study of the broadcast use case, we believe our analysis is applicable to many other classes of audio-video use cases.

This approach holds promise for achieving interoperability and access at scale, but practical progress depends on production and presentation deployment that are mutually dependent. At the time of writing, governmental review of available technical paths is being undertaken in some geographies with some urgency, providing those who take initiative with an opportunity to exert substantial influence on the path taken.

## REFERENCES

1. ATSC A/334:2024. Audio Watermark Emission. Advanced Television Systems Committee. https://www.atsc.org/atsc-documents/a3342016-audio-watermark-emission/.

2. ATSC A/335:2024. Video Watermark Emission. Advanced Television Systems Committee. https://www.atsc.org/atsc-documents/a3352016-video-watermark-emission/.

3. ATSC A/336:2024. Content Recovery in Redistribution Scenarios. Advanced Television Systems Committee. https://www.atsc.org/atsc-documents/a3362017-content-recovery-redistribution-scenarios/.

4. C2PA Specifications v2.0. 2024. Coalition for Content Provenance and Authenticity. https://c2pa.org/specifications/specifications/2.0/index.html.

5. Parsons, A. 2024. Durable Content Credentials. Content Authenticity Initiative Blog. https://contentauthenticity.org/blog/durable-content-credentials.

6. ETSI TS 126 346 v13.3.0 (2016-01). Universal Mobile Telecommunications Systems (UMTS); LTE; Multimedia Broadcast/Multicast Service (MBMS); Protocols and codecs (3GPP TS 26.346 version 13.3.0 Release 13). European Telecommunications Standards Institute. http://www.etsi.org/deliver/etsi_ts/126300_126399/126346/13.03.00_60/ts_126346v130300p.pdf.

7. Targeted Advertising Verification and Validation Test Streams – Watermarking. 2023. DVB. https://dvb.org/specifications/verification-validation/targeted-advertising-watermarking/.

8. Recommendation ITU-R BS.1116-3: Methods for the Subjective Assessment of Small Impairments in Audio Systems including Multi-channel Sound Systems. 2014. ITU Radio Communication Assembly.

9. ATSC A/339:2024. Audio Watermark Modification and Erasure. Advanced Television Systems Committee. https://www.atsc.org/atsc-documents/3392017-atsc-recommended-practice-audio-watermark-modification-erasure/.

10. Zhao, X., et al. Invisible Image Watermarks Are Provably Removable Using Generative AI. Preprint. https://doi.org/10.48550/arXiv.2306.01953.

11. Kirwan, C., et al. Repetition of Computer Security Warnings Results in Differential Repetition Suppression Effects as Revealed With Functional MRI. 2020. Frontiers in Psychology. Vol 11, 2020.  https://doi.org/10.3389/fpsyg.2020.528079.

12. England, P. et al. 2021. AMP: Authentication of Media via Provenance. 12th ACM Multimedia Systems Conference. July 2021.    https://doi.org/10.48550/arXiv.2001.07886.

13. SMPTE 2064-1:2015. Fingerprint Generation. Society of Motion Picture and Television Engineers. https://ieeexplore.ieee.org/document/7395520.

14. ISO/IEC 23001-7:2006, "Information technology – MPEG systems technologies – Part 7: Common Encryption in ISO base media file format files", first edition.

15. ISO/IEC 23000-19:2020, "Information technology – Multimedia application format (MPEG-A) – Part 19: Common media application format (CMAF) for segmented media".

16. C2PA Explainer, release 1.4, 2023, https://c2pa.org/specifications/specifications/1.3/explainer/Explainer.html

17. C2PA Implementation Guidance, release 1.4, 2023, https://c2pa.org/specifications/specifications/1.3/guidance/Guidance.html

18. C2PA User Experience Guidance for Implementers, release 1.0, 2023, https://c2pa.org/specifications/specifications/1.4/ux/UX_Recommendations.html

19. C2PA Security Considerations, release 1.0, 2023, https://c2pa.org/specifications/specifications/1.0/security/Security_Considerations.html

20. C2PA Harms Modeling, release 1.0, 2023, https://c2pa.org/specifications/specifications/1.0/security/Harms_Modelling.html

21. C2PA Guidance for Artificial Intelligence and Machine Learning, release 1.3, 2023, https://c2pa.org/specifications/specifications/1.3/ai-ml/ai_ml.html

22. ISO/IEC 19566-5:2023, Information technologies, JPEG systems, Part 5: JPEG universal metadata box format (JUMBF)

23. RFC 2387, 1998, The MIME Multipart/Related Content-type, Internet Engineering Task Force. https://datatracker.ietf.org/doc/html/rfc2387.

24. RFC 6838, 2013, Media Type Specifications and Registration Procedures, Internet Engineering Task Force. https://datatracker.ietf.org/doc/html/rfc6838.

25. ISO/IEC ISO 23009-1:2022, Information technology – Dynamic adaptive streaming over HTTP (DASH) – Part 1: Media presentation description and segment formats